\documentclass[sigconf]{acmart}

\usepackage{multirow}
\usepackage{colortbl}
\usepackage{algorithm}
\usepackage{algpseudocode}
\usepackage{booktabs}
\usepackage{array}
\usepackage{tabularx}
\usepackage{amsmath,amsthm,mathtools}
\usepackage{subcaption}
\usepackage{xcolor}
\usepackage{siunitx}

\definecolor{phbg}{HTML}{FFF3CD}
\definecolor{phfg}{HTML}{7C4700}

\AtBeginDocument{}

\acmConference[ASP-DAC'27]{32nd Asia and South Pacific Design Automation Conference}{Jan. 25-28, 2027}{Tokyo, Japan}

\begin{document}

\title{NOVA-CIM: Noise- and Correlation-Tolerant Stochastic Interfaces for Analog Compute-in-Memory}

\author{Jiachen Ren$^{1,\dagger}$, Wenshuai Yao$^{2,\dagger}$, Haobo Liu$^1$, Xincheng Feng$^1$, Chenxi Hu$^1$, Zhengwu Liu$^1$, \\  Kechao Tang$^2$, Wenyong Zhou$^{1,*}$, Ngai Wong$^{1,*}$}
\authornote{$^{\dagger}$: Equal contribution. $^{*}$: Corresponding authors.}
\affiliation[obeypunctuation=true]{%
  \institution{$^1$Department of Electrical and Computer Engineering, The University of Hong Kong}
  \city{, Hong Kong}
  \country{, China}%
}
\affiliation[obeypunctuation=true]{%
  \institution{$^2$School of Integrated Circuits, Peking University}
  \city{, Beijing}
  \country{, China}%
}

\begin{abstract}
Analog compute-in-memory (CIM) enables energy-efficient model acceleration, but its reliance on ADC-based readout, which directly quantizes noisy column currents, makes inference accuracy highly sensitive to analog read noise, active-row scaling, and ADC precision.
In this paper, we present \textbf{NOVA-CIM}, a noise- and correlation-tolerant stochastic interface for analog CIM by replacing multi-bit ADC readout with random-reference 1-bit sensing and reconstructing results through lightweight counting. By converting column currents into comparison probabilities, this probability-domain readout averages zero-mean dynamic read noise over stochastic samples while reducing dependence on high-resolution ADCs. We provide a unified robustness analysis showing that dynamic read noise is suppressed through temporal averaging and that spatial input-bitstream correlation increases instantaneous current variance rather than introducing first-order MAC bias.
MAC-level experiments and end-to-end evaluation on ViT-Base validate the analysis: under read noise, Top-1 accuracy remains $84.48\%$ near the $84.51\%$ bfloat16 (BF16) baseline; under stochastic number generator (SNG) reuse, MAC bias stays near zero while root-mean-square error (RMSE) and stochastic cross-correlation (SCC) grow as predicted.
\end{abstract}

\ccsdesc[500]{Hardware~Hardware accelerators}

\keywords{Compute-in-memory, stochastic computing, stochastic interface, neural network acceleration}

\maketitle

\section{Introduction}

Deep neural network inference is dominated by matrix-vector multiplication. Conventional architectures pay large energy and latency for moving weights and activations between memory and compute. Compute-in-memory (CIM) performs computation inside memory arrays to cut data movement, and analog CIM executes matrix-vector multiplication through analog current or charge accumulation with high parallelism~\cite{shafiee2016isaac,chi2016prime,le202364,wan2022compute}.

Analog CIM is ultimately limited by the mixed-signal interface to digital logic. Digital activations drive the array, weights set conductance, and column current carries the multiply-accumulate (MAC) result, requiring digital-to-analog converters (DACs) for input and analog-to-digital converters (ADCs) for output readout. Deterministic conversion dominates area, power, and latency, and array scaling is often limited by per-column readout and per-row encoding rather than in-array computation~\cite{singh2025design,yu202265,mu2022sram}. Multi-bit ADC readout maps analog perturbations directly into digital errors amid programming error, read noise, sensing variation, reference fluctuation, and row-dependent accumulation noise~\cite{zhang2025asim,antolini2023combined}. Quantization encodes both signal and instantaneous noise into a fixed digital word, so accuracy depends on ADC resolution, noise, variation, and accumulated current range, which grows with parallel row activation. Higher ADC resolution improves fidelity but further increases peripheral cost~\cite{zhang2023macc,yu202265,wu202322nm}.

A readout interface is needed that preserves array parallelism without scaling converter cost proportionally. Stochastic computing (SC) encodes values by the probability of ones in a bitstream~\cite{alaghi2013survey} and supports estimation by temporal averaging. For analog CIM, comparing column current against a random reference over multiple cycles yields 1-bit samples whose mean estimates the normalized MAC, replacing multi-bit ADC readout with 1-bit sensing and counting~\cite{rogers2024stox}. Stochastic input encoders may share linear feedback shift registers (LFSRs) across rows to reduce overhead~\cite{hsiao2021zero,baker2020hypergeometric}. Logic SC treats such dependence as harmful for AND multiplication~\cite{alaghi2013exploiting,asadi2021corld}, but analog CIM performs conductance-weighted accumulation with stochastic bits gating deterministic weights, so input-side dependence does not alter the first-order MAC expectation.

\begin{figure}[t]
  \centering
  \includegraphics[width=\linewidth]{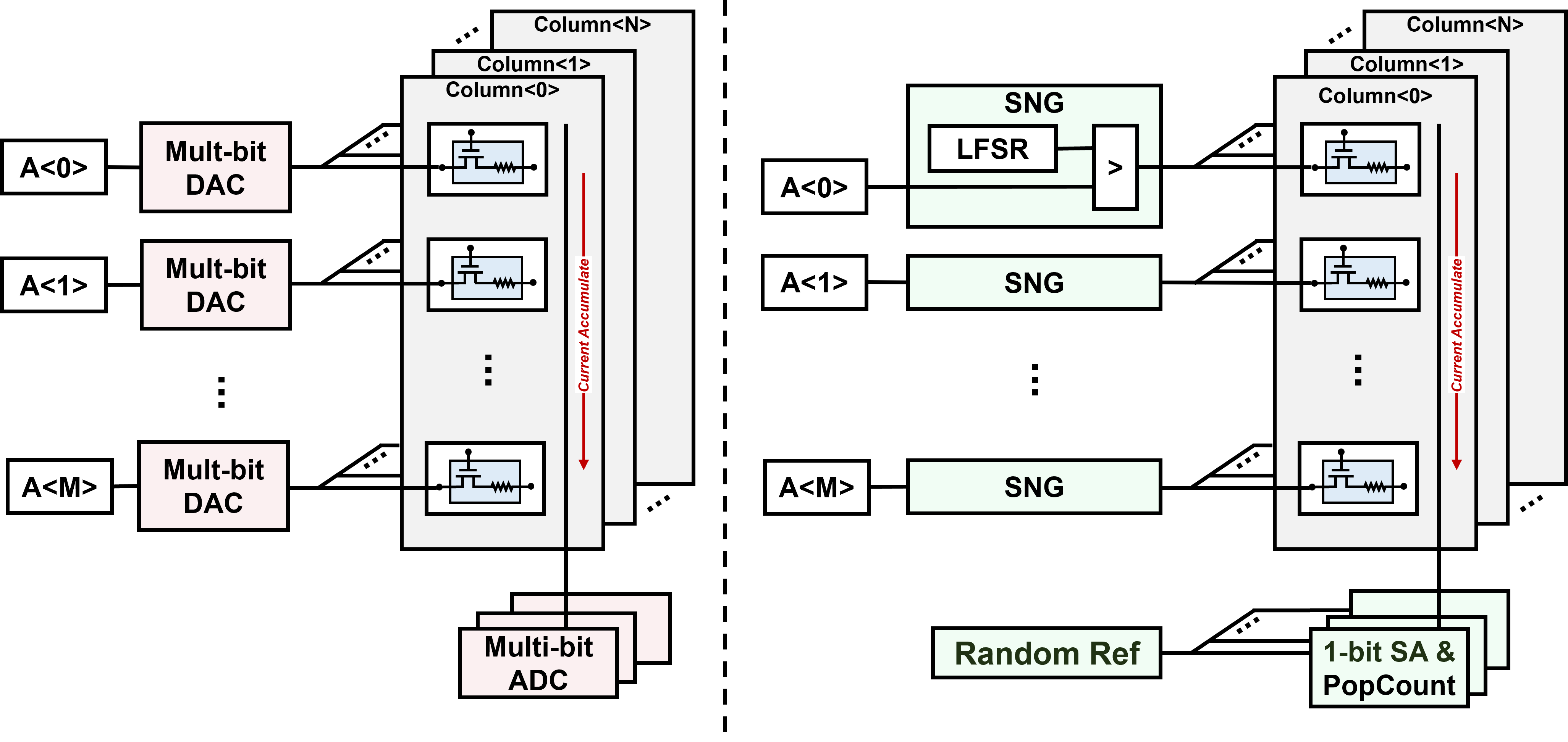}
  \caption{Overview of conventional ADC-based analog CIM readout and the proposed NOVA-CIM stochastic interface.}
  \label{fig:nova_arch}
\end{figure}

We present \textbf{NOVA-CIM}, a stochastic interface for analog CIM that reconstructs column MAC results by random-reference one-bit sensing and temporal counting instead of multi-bit ADC quantization, as illustrated in Figure~\ref{fig:nova_arch}. For this probability-domain readout, we show that the normalized MAC estimator is unbiased with variance $O(1/N)$, that zero-mean dynamic read noise introduces no first-order bias and is attenuated by temporal averaging, and that spatial input correlation under conductance-weighted accumulation raises instantaneous current variance without biasing the MAC expectation. Experiments at MAC and network levels validate noise robustness and correlation-tolerant estimation under shared stochastic number generators (SNGs).

The main contributions of this paper are:
\begin{itemize}
    \item We formulate NOVA-CIM as a probability-domain CIM readout based on random-reference one-bit sensing and temporal counting, and prove that the normalized MAC estimator is unbiased with variance $O(1/N)$.
    \item We show that zero-mean dynamic read noise introduces no first-order bias and is attenuated by $1/N$ averaging, unlike single-shot multi-bit ADC quantization.
    \item We prove that spatial input correlation does not bias the MAC expectation under conductance-weighted accumulation and only increases instantaneous current variance, unlike correlation-induced bias in logic stochastic multiplication.
    \item Experiments confirm both properties: end-to-end evaluation on ViT-Base retains $84.48\%$ Top-1 under read noise near the $84.51\%$ bfloat16 (BF16) baseline, and SNG reuse keeps MAC bias near zero while root-mean-square error (RMSE) and stochastic cross-correlation (SCC) grow as predicted.
\end{itemize}

\section{Related Work}

\subsection{Analog CIM and Data-Conversion Interfaces}

Analog CIM arrays in SRAM, ReRAM, PCM, and related technologies accumulate analog current on bit-lines with high parallelism~\cite{shafiee2016isaac,wang2024rescim,le202364,wan2022compute}, yet scaling is often limited by per-column readout and per-row activation periphery rather than in-array computation~\cite{zhang2025asim}. Most reported macros therefore rely on multi-bit ADCs or comparably complex sense paths.

Prior work lowers converter cost through shared or lower-resolution ADCs, time-multiplexed sensing, and hierarchical digitization~\cite{chi2016prime,yu202265,yuan202434,wu202322nm}. Each readout event still yields a deterministic multi-bit word, so noise and programming error are quantized with the MAC result and accuracy remains coupled to ADC resolution and sensing bandwidth. Algorithm-level quantization and calibration~\cite{jacob2018quantization,lammie2024improving,zhou2025distribution,zhou2025binary} do not alter this readout mapping.

Probability-domain interfaces compare column current to stochastic references over multiple cycles and reconstruct MAC values by digital accumulation rather than multi-bit quantization. StoX-Net applies this at the column readout~\cite{rogers2024stox}; memristive designs process unary bitstreams inside the crossbar~\cite{alam2022stochastic}. These works preserve inference accuracy under independent encoders but do not analyze row-wise sharing of linear feedback shift registers under conductance-weighted accumulation~\cite{hoffend2022analyzing}.

\subsection{Stochastic Bitstream Dependence}

Stochastic dependence between bitstreams has been studied extensively in logic-based stochastic computing, where accuracy is operator specific. Alaghi and Hayes~\cite{alaghi2013exploiting} formalized SCC between two streams and showed that AND multiplication requires near-zero dependence, whereas MIN and MAX operators can benefit from positive dependence. These results address pairwise logic operations on stochastic operands. Analog in-memory MAC instead forms a conductance-weighted sum in which stochastic input bits gate deterministic weights, so the first-order result depends on input marginal probabilities rather than on products of distinct input streams.

Shared LFSRs further couple the random bits delivered to multiple rows. Hsiao et al.~\cite{hsiao2021zero} proposed zero correlation error (ZCE) to measure how closely two finite-length streams approximate independence, and Baker and Hayes~\cite{baker2020hypergeometric} showed that Bernoulli models can misestimate error when the random source has finite internal state. These metrics are useful for reporting bitstream statistics under row-wise SNG reuse, but they were not developed to predict MAC accuracy in analog CIM. We therefore use SCC and ZCE only as characterization tools in experiments, while MAC correctness is established by the accumulation analysis and by measured MAC and inference accuracy.

\subsection{Positioning of NOVA-CIM}

\begin{table}[!t]
  \centering
  \caption{Comparison of readout interfaces for analog CIM.}
  \label{tab:positioning}
  \small
  \setlength{\tabcolsep}{5pt}
  \renewcommand{\arraystretch}{1.45}
  \begin{tabular}{@{}lccc@{}}
    \toprule
    \textbf{Interface} & \textbf{Readout} & \textbf{MAC} & \textbf{SNG reuse} \\
    \midrule
    ADC-CIM   & Multi-bit ADC       & Fixed-$G$ accum. & --- \\
    Logic SC  & Bitstream logic     & Logic products & AND bias \\
    NOVA-CIM  & 1-bit ref.\ + count & Input gating   & No MAC bias \\
    \bottomrule
  \end{tabular}
\end{table}

Table~\ref{tab:positioning} summarizes the distinction. Prior analog CIM macros optimize the mixed-signal boundary by reducing ADC resolution, sharing converters, or tightening sensing schedules~\cite{chi2016prime,yu202265,singh2025design,wan2022compute}; NOVA-CIM instead replaces deterministic multi-bit quantization of column currents with probability-domain estimation using random-reference 1-bit sensing and popcount reconstruction. Conductance programming and analog accumulation remain unchanged, while precision is controlled by bitstream length and temporal averaging rather than ADC resolution, allowing zero-mean noise to average out across stochastic cycles. Unlike logic-based SC, NOVA-CIM does not multiply stochastic operand streams: stochastic inputs only gate deterministic conductance.

\begin{figure}[!t]
\centering
\includegraphics[width=\linewidth]{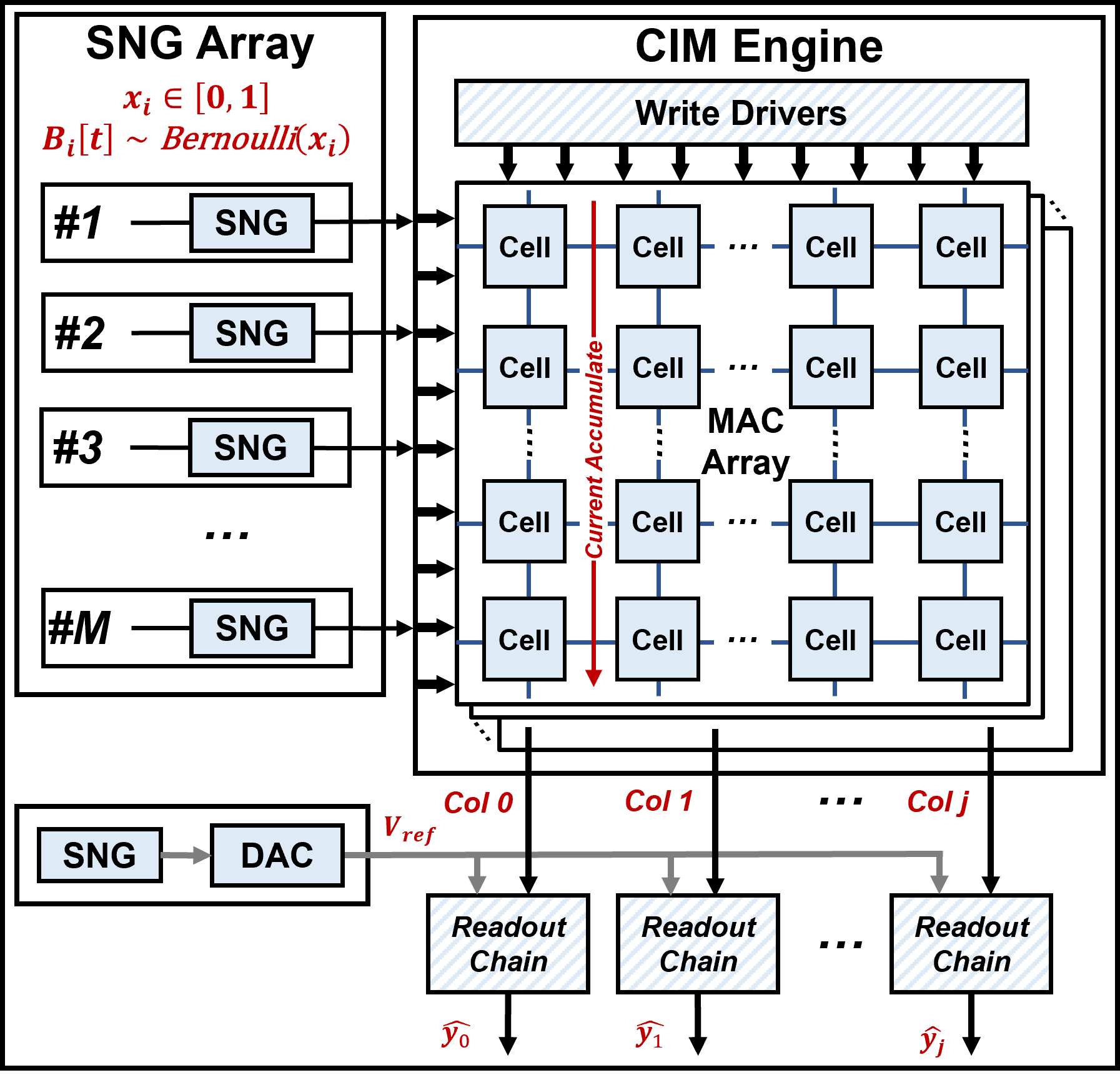}
\caption{NOVA-CIM macro-level vector-MAC dataflow: per-row SNGs generate Bernoulli activation streams, and column currents are reconstructed through the readout chain.}
\label{fig:micro_architecture}
\vspace{-0.4cm}
\end{figure}

\section{NOVA-CIM Architecture}

\begin{figure}[!t]
\centering
\includegraphics[width=\linewidth]{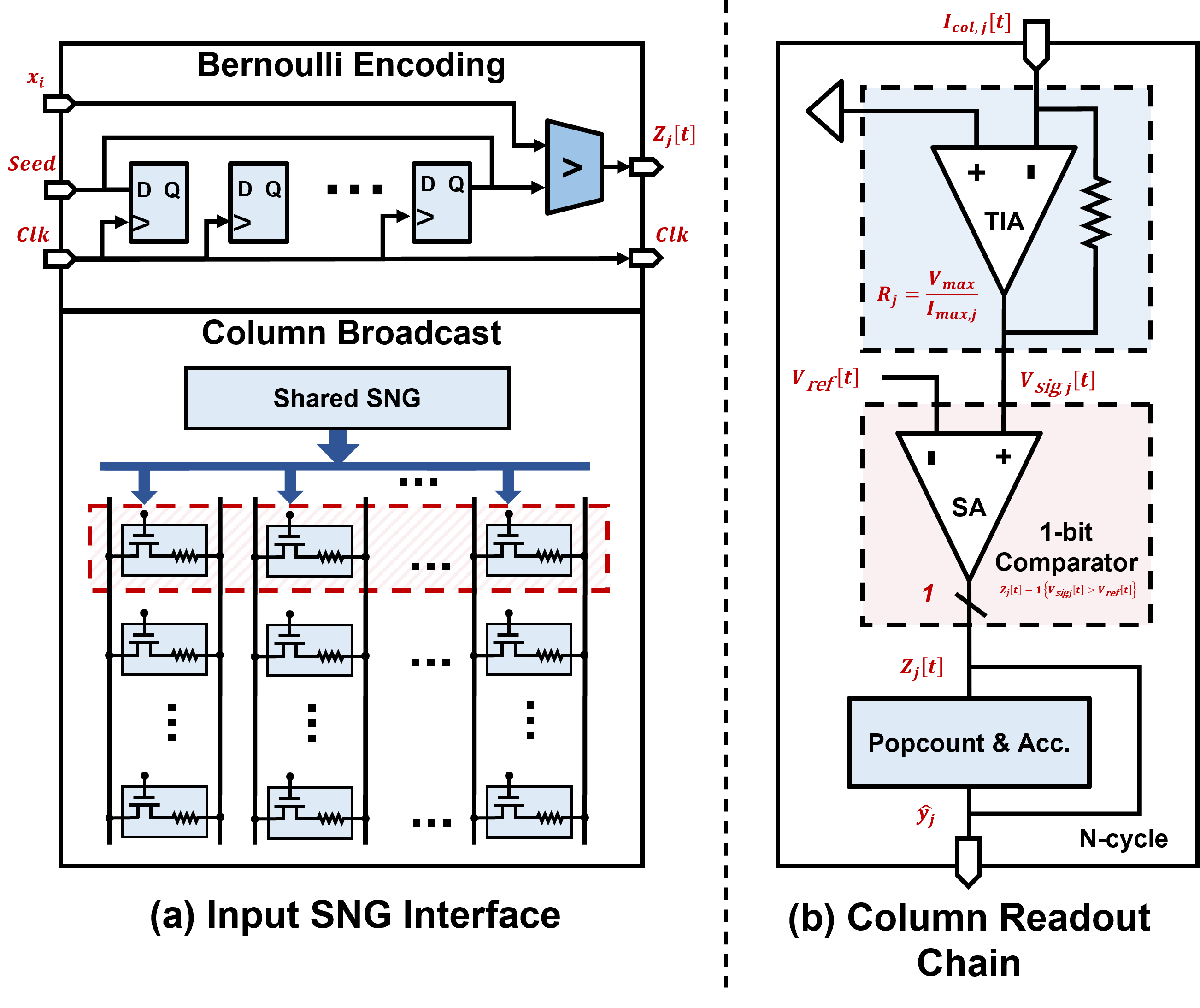}
\caption{NOVA-CIM stochastic interface. (a) Bernoulli input encoding and column broadcast. (b) Per-column random-reference readout with temporal accumulation over $N$ cycles.}
\label{fig:nova_interface}
\vspace{-0.4cm}
\end{figure}

Figure~\ref{fig:micro_architecture} shows the NOVA-CIM organization for one VMAC. A bank of $M$ per-row SNGs maps normalized activations $x_i$ to Bernoulli streams $B_i[t]$ that drive the word lines of an $M\times C$ MAC array. Write drivers program cell conductances, and each column accumulates a Kirchhoff current in parallel. A shared SNG and DAC generate the randomized reference $V_{\mathrm{ref}}$ for all column readout chains. Over $N$ stochastic cycles, each chain reconstructs $\hat{y}_j$ by random-reference 1-bit sensing and popcount. Figure~\ref{fig:nova_interface} details Bernoulli encoding with column broadcast and the column readout from transimpedance-amplifier (TIA) conversion through 1-bit comparison to $N$-cycle accumulation.

\subsection{Stochastic Interface Model}

Activations are mapped to $x_i \in [0,1]$ before Bernoulli encoding; signed inputs enter through an affine map into this range. The SNG yields $B_i[t] \sim \mathrm{Bernoulli}(x_i)$ and drives word-line $i$ with
\begin{equation}
    V_i[t] = V_{DD} B_i[t],
\end{equation}
so each row receives a 1-bit drive rather than a multi-bit analog level. An LFSR and a comparator on $x_i$ form the encoder.

Weights are programmed as conductances $G_{i,j}$. At cycle $t$, bit-line $j$ carries
\begin{equation}
    I_{\mathrm{col},j}[t] = V_{DD}\sum_{i=1}^{M} G_{i,j} B_i[t].
\end{equation}
Stochastic bits gate row injection while $G_{i,j}$ remain fixed; the in-array operation is a weighted sum, not a product of stochastic streams.

With $I_{\max,j} = V_{DD}\sum_{i} G_{i,j}$, cycle $t$ draws $I_{\mathrm{ref},j}[t]$ uniformly from $0$ to $I_{\max,j}$ and sets
\begin{equation}
    Z_j[t] = \mathbf{1}\{I_{\mathrm{col},j}[t] > I_{\mathrm{ref},j}[t]\}.
\end{equation}
After $N$ cycles,
\begin{equation}
    \hat{p}_j = \frac{1}{N}\sum_{t=1}^{N} Z_j[t], \qquad
    \hat{y}_j = y_{\max,j}\hat{p}_j,
\end{equation}
where $y_{\max,j}=\sum_{i} G_{i,j}$. In hardware, a TIA with $R_j=V_{\max}/I_{\max,j}$ converts $I_{\mathrm{col},j}[t]$ to $V_{\mathrm{sig},j}[t]$, the shared DAC supplies $V_{\mathrm{ref}}[t]$, and a 1-bit sense amplifier (SA) resolves the equivalent comparison $Z_j[t]=\mathbf{1}\{V_{\mathrm{sig},j}[t]>V_{\mathrm{ref}}[t]\}$.

\subsection{Macro Organization and Design Choices}

For VMAC, a single encoder bank generates $B_{\mathrm{vec}}[t]$ and broadcasts it to all $C$ columns. The baseline assigns one LFSR to each row. Row-clustered reuse shares one LFSR within each row block to reduce encoder area, which increases spatial bitstream correlation across rows. Such correlation increases instantaneous current variance but leaves the first-order MAC expectation unchanged.

Each column integrates a TIA, a 1-bit SA, a reference sampler, and a popcount accumulator. One reference waveform $V_{\mathrm{ref}}[t]$ may be broadcast to all columns when transimpedance gain satisfies $R_j = V_{\max}/I_{\max,j}$ at each column full scale $I_{\max,j}$. A per-column reference generator remains available at higher peripheral cost. Readout precision is set by the bitstream length $N$ rather than by multi-bit converter resolution.

One output vector is reconstructed over $N$ stochastic cycles. During cycle $t$, the encoder emits $B_{\mathrm{vec}}[t]$, the crossbar forms $I_{\mathrm{col},j}[t]$ on every column, the reference sampler draws $I_{\mathrm{ref},j}[t]$, the comparator resolves $Z_j[t]$, and the digital unit updates each popcount. Readout latency therefore scales linearly with $N$.

Conductance programming and Kirchhoff current accumulation follow the same array organization as ADC-based CIM. Cell weights remain stored as conductances, and column currents still form the analog MAC. NOVA-CIM changes only the mixed-signal interface that maps each column current to a digital MAC estimate through random-reference one-bit sensing and temporal counting.

\section{Robustness Analysis}

This section analyzes the robustness of NOVA-CIM, showing that random-reference readout provides an unbiased estimate of the normalized MAC result, zero-mean dynamic read noise is averaged over stochastic samples, and spatial input-bitstream correlation introduces no first-order MAC bias.

\subsection{Ideal Stochastic Estimation}
\label{sec:ideal_stochastic_estimation}

Consider one nonnegative CIM column with \(M\) active rows. Signed weights can be represented using differential columns, to which the same analysis applies separately. Let \(x_i\in[0,1]\) be the normalized input activation and \(G_i\geq 0\) be the programmed conductance. The ideal MAC and its full-scale value are
\begin{equation}
    y=\sum_{i=1}^{M}x_iG_i,
    \qquad
    y_{\max}=\sum_{i=1}^{M}G_i,
\end{equation}
and the normalized MAC is
\begin{equation}
    p=\frac{y}{y_{\max}}.
\end{equation}

At stochastic cycle \(t\), the input is represented by binary variables \(B_i[t]\in\{0,1\}\) satisfying
\begin{equation}
    \mathbb{E}[B_i[t]]=x_i.
\end{equation}
The instantaneous column current is
\begin{equation}
    I[t]=V_{DD}\sum_{i=1}^{M}G_iB_i[t],
    \qquad
    I_{\max}=V_{DD}\sum_{i=1}^{M}G_i.
\end{equation}
Since \(0\leq I[t]\leq I_{\max}\), comparing \(I[t]\) with an independent random reference
\begin{equation}
    I_{\mathrm{ref}}[t]\sim\mathcal{U}(0,I_{\max}).
\end{equation}
produces
\begin{equation}
    Z[t]=\mathbf{1}\{I[t] > I_{\mathrm{ref}}[t]\}.
\end{equation}
Conditioned on \(I[t]\),
\begin{equation}
    \Pr(Z[t]=1\mid I[t])=\frac{I[t]}{I_{\max}}.
\end{equation}
Therefore,
\begin{equation}
    \mathbb{E}[Z[t]]
    =
    \frac{\mathbb{E}[I[t]]}{I_{\max}}
    =
    \frac{\sum_{i=1}^{M}G_i\mathbb{E}[B_i[t]]}{\sum_{i=1}^{M}G_i}
    =
    \frac{\sum_{i=1}^{M}G_i x_i}{\sum_{i=1}^{M}G_i}
    =
    p.
\end{equation}
Thus,
\begin{equation}
    \hat{p}=\frac{1}{N}\sum_{t=1}^{N}Z[t],
\end{equation}
is an unbiased estimator of the normalized MAC:
\begin{equation}
    \mathbb{E}[\hat{p}]=p.
\end{equation}
For temporally independent output samples,
\begin{equation}
    \mathrm{Var}(\hat{p})=\frac{p(1-p)}{N}.
\end{equation}

\subsection{Dynamic Read Noise Averaging}
\label{sec:dynamic_read_noise_averaging}

With dynamic read noise, the sensed current becomes
\begin{equation}
    \tilde{I}[t]=I[t]+n_r[t],
\end{equation}
where \(n_r[t]\) is zero-mean and independent across cycles. The readout bit is
\begin{equation}
    Z[t]=\mathbf{1}\{\tilde{I}[t] > I_{\mathrm{ref}}[t]\}.
\end{equation}
Ignoring boundary clipping, or equivalently assuming sufficient reference guard margin,
\begin{equation}
    \Pr(Z[t]=1\mid I[t],n_r[t])
    =
    \frac{I[t]+n_r[t]}{I_{\max}}.
\end{equation}
Taking expectation over \(n_r[t]\) gives
\begin{equation}
    \mathbb{E}[Z[t]\mid I[t]]
    =
    \frac{I[t]}{I_{\max}}
    +
    \frac{\mathbb{E}[n_r[t]]}{I_{\max}}
    =
    \frac{I[t]}{I_{\max}}.
\end{equation}
Therefore, zero-mean dynamic read noise does not introduce first-order bias.

The normalized noise component is averaged over \(N\) cycles:
\begin{equation}
    \mathrm{Var}
    \left(
    \frac{1}{N}\sum_{t=1}^{N}\frac{n_r[t]}{I_{\max}}
    \right)
    =
    \frac{\sigma_r^2}{N I_{\max}^2},
\end{equation}
where \(\sigma_r^2=\mathrm{Var}(n_r[t])\). Thus, temporally independent read perturbations are suppressed by stochastic averaging, while the intrinsic binary sampling fluctuation also scales as \(1/N\).

\subsection{Correlation-Tolerant MAC Estimation}
\label{sec:correlation_tolerant_estimation}

In logic-based stochastic computing, multiplying two stochastic streams \(A[t]\) and \(W[t]\) gives
\begin{equation}
    \mathbb{E}[A[t]W[t]]
    =
    \mathbb{E}[A[t]]\mathbb{E}[W[t]]
    +
    \mathrm{Cov}(A[t],W[t]),
\end{equation}
so operand correlation directly creates bias.

NOVA-CIM avoids this first-order correlation bias because stochastic input bits gate deterministic conductances:
\begin{equation}
    I[t]=V_{DD}\sum_{i=1}^{M}G_iB_i[t].
\end{equation}
Even if \(B_i[t]\) and \(B_j[t]\) are spatially correlated,
\begin{equation}
    \mathbb{E}[I[t]]
    =
    V_{DD}\sum_{i=1}^{M}G_i\mathbb{E}[B_i[t]]
    =
    V_{DD}\sum_{i=1}^{M}G_ix_i.
\end{equation}
Since the random reference is independent,
\begin{equation}
    \mathbb{E}[Z[t]]
    =
    \frac{\mathbb{E}[I[t]]}{I_{\max}}
    =
    p.
\end{equation}
Thus, spatial input-bitstream correlation does not introduce first-order MAC bias.
The instantaneous-current variance is
\begin{equation}
    \mathrm{Var}(I[t])
    =
    V_{DD}^2
    \Bigg(
    \sum_{i=1}^{M}G_i^2\mathrm{Var}(B_i[t])
    +
    2\sum_{i<j}G_iG_j
    \mathrm{Cov}(B_i[t],B_j[t])
    \Bigg).
\end{equation}
Spatial covariance therefore increases $\mathrm{Var}(I[t])$ without changing $\mathbb{E}[Z[t]]$ or $\mathbb{E}[\hat{p}]$.


\section{Experiments}

\subsection{Experiment Setup}

We evaluate NOVA-CIM along three primary axes: estimator correctness under temporal averaging, robustness to read noise and active-row scaling at network level, and correlation tolerance under SNG reuse. Interface area and digital datapath cost are reported at the end.

At the MAC level, Monte Carlo simulation evaluates the normalized target $p = y/y_{\max}$ over $16\leq N\leq4096$. We report bias, mean-squared error (MSE) or RMSE, and SCC where applicable, and compare against Shared-SNG SC, which is sensitive to correlation. Encoder-reuse measurements use the proposed input-SNG organization with a 12-bit random-reference DAC and unsigned Q0.16 operands.

At the network level, programming noise is modeled as a static multiplicative conductance perturbation, whereas read noise is zero-mean, independent Gaussian current noise normalized to $I_{\max}$. We compare NOVA-CIM with ADC-CIM at finite ADC resolutions on ViT-Base using ImageNet-1k and on Pythia-160M and Pythia-410M using one million C4 tokens. The network evaluation uses expectations in the probability domain to isolate analog nonidealities from sampling variance due to finite bitstreams.

For implementation cost, ADC-CIM and NOVA-CIM interfaces are evaluated at the same 65\,nm process node, and field-programmable gate array (FPGA) post-route results report the digital datapath.

\subsection{Experiment Results}
\begin{figure}[!t]
\centering
\includegraphics[width=\columnwidth]{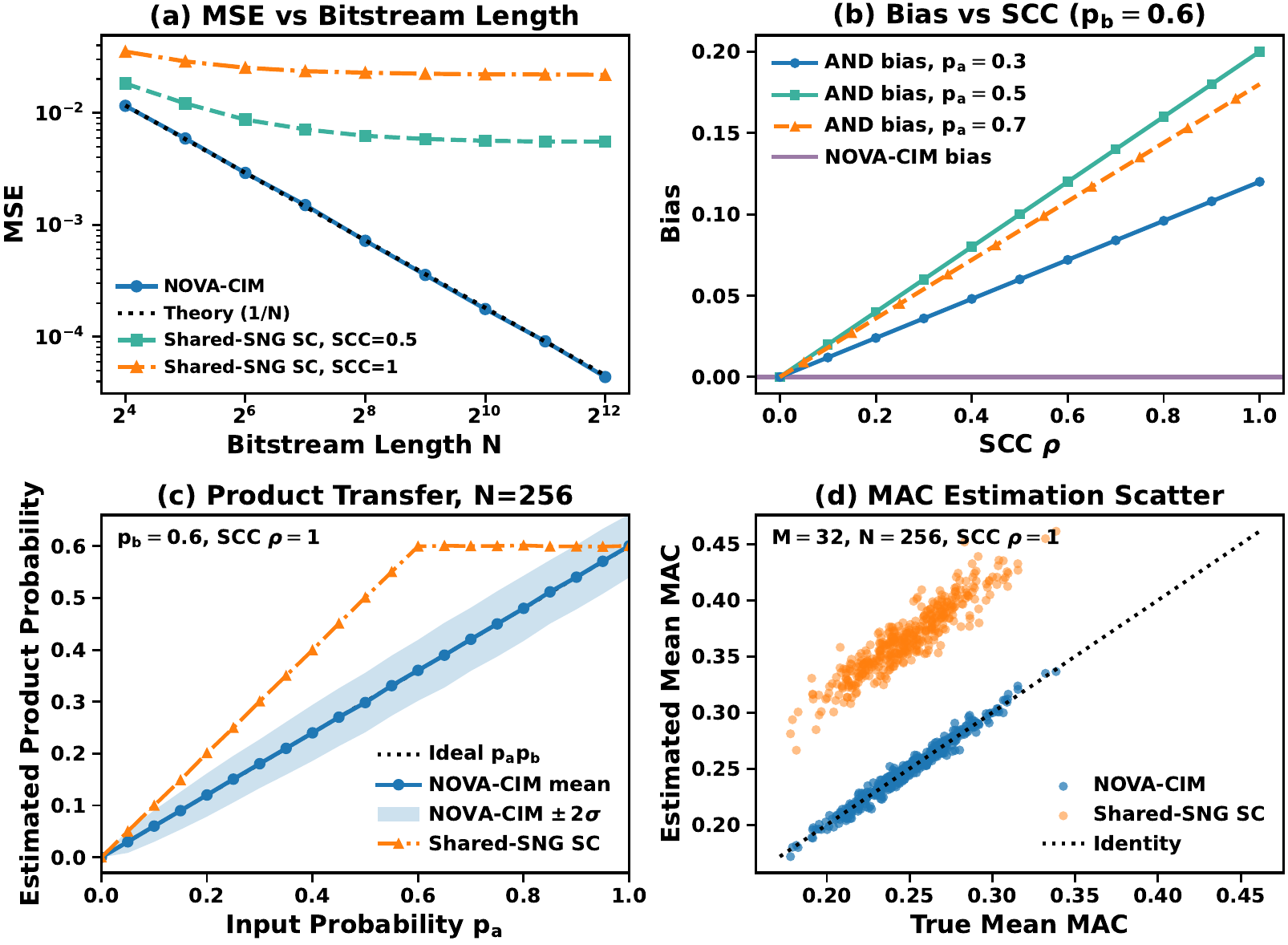}
\caption{Estimator correctness evaluation. (a) MSE versus bitstream length. (b) Bias versus operand SCC at $p_b=0.6$. (c) Product transfer at $N=256$. (d) MAC estimation scatter for $M=32$ and $N=256$.}
\label{fig:cat1}
\end{figure}

\textbf{Estimator Correctness.}
Consistent with the unbiased estimator and $O(1/N)$ variance derived in Section~\ref{sec:ideal_stochastic_estimation}, Figure~\ref{fig:cat1}(a) shows that the NOVA-CIM MSE decreases from $1.15\times10^{-2}$ at $N=16$ to $4.38\times10^{-5}$ at $N=4096$, whereas Shared-SNG SC retains SCC-dependent error floors. Figure~\ref{fig:cat1}(b) further shows that operand SCC directly biases stochastic AND multiplication. In Figures~\ref{fig:cat1}(c) and (d), NOVA-CIM follows the ideal product and MAC identity, whereas Shared-SNG SC exhibits systematic positive bias.

\begin{figure}[!t]
\centering
\includegraphics[width=\columnwidth]{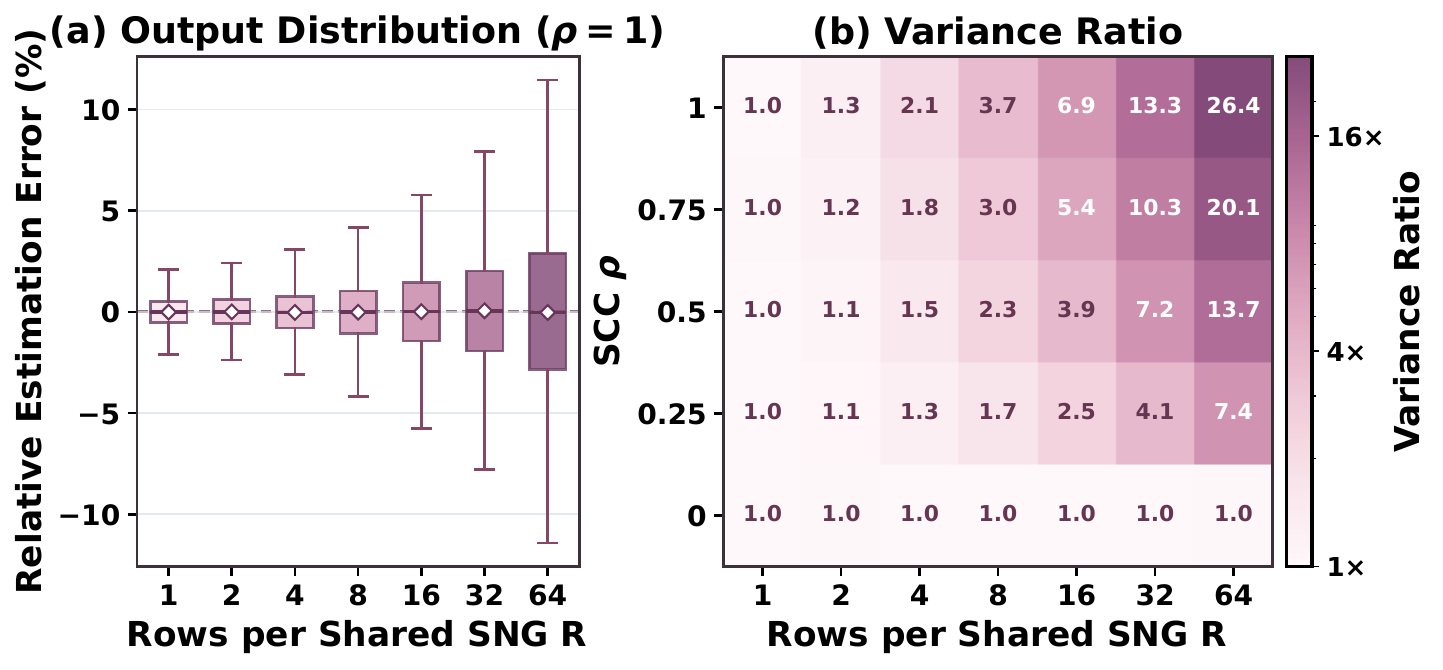}
\caption{Controlled row correlation experiment at $M=64$ and $N=256$. (a) Relative error distributions at $\rho=1$ versus rows per shared SNG $R$. (b) Variance ratio sweep over $R$ and $\rho$; colors show Monte Carlo results and cell values show theory.}
\label{fig:e001_row_correlation_robustness}
\end{figure}

\textbf{Correlation Tolerance.}
Unlike operand correlation in stochastic AND multiplication, Section~\ref{sec:correlation_tolerant_estimation} predicts that row-wise SNG sharing in NOVA-CIM preserves every Bernoulli marginal and therefore the expected MAC. Figure~\ref{fig:e001_row_correlation_robustness} validates this property for $M=64$ and $N=256$. The output distributions in Figure~\ref{fig:e001_row_correlation_robustness}(a) remain centered near zero but widen as more rows share an SNG. Figure~\ref{fig:e001_row_correlation_robustness}(b) shows that the variance of the instantaneous column current increases with both $R$ and $\rho$; full 64-row sharing at $\rho=1$ produces a $26.7\times$ variance ratio, close to the predicted $26.4\times$. Row correlation therefore widens the instantaneous current distribution without shifting the first-order mean.

Table~\ref{tab:sng_reuse} reports the corresponding area-accuracy trade-off for the proposed input-SNG microarchitecture at $M{=}64$ and $N{=}1024$. Mean absolute SCC rises from $0.08$ for per-row encoding to $0.98$ for one-bit broadcast, while MAC bias stays near zero for all methods. RMSE grows with sharing strength. Clustered encoding with $K{=}8$ reduces input-SNG look-up table (LUT) count from $2537$ to $745$, equal to $0.29\times$ the per-row baseline, and raises RMSE only from $0.653$ to $0.678$. Encoder reuse is thus an area-accuracy trade-off for NOVA-CIM: bitstream dependence and variance increase, but the first-order MAC expectation does not.

\begin{table}[!t]
\centering
\caption{Input SNG reuse under the proposed encoder organization at $M{=}64$ and $N{=}1024$.}
\label{tab:sng_reuse}
\renewcommand{\arraystretch}{1.08}
\begin{tabular}{@{}lccccc@{}}
\toprule
\textbf{Method} & \textbf{\#LFSR} & \textbf{$|\mathrm{SCC}|$} & \textbf{Bias} & \textbf{RMSE} & \textbf{LUT} \\
\midrule
Per-row             & 64 & 0.08 & $-0.085$ & \textbf{0.653} & 2537 \\
Clustered, $K{=}16$ & 16 & 0.10 & $-0.079$ & 0.664 & 1005 \\
Clustered, $K{=}8$  &  8 & 0.16 & $-0.079$ & 0.678 & \textbf{745} \\
Broadcast           &  1 & 0.98 & $-0.050$ & 0.816 & 523 \\
\bottomrule
\end{tabular}
\vspace{-0.4cm}
\end{table}

\begin{figure}[!t]
\centering
\includegraphics[width=\columnwidth]{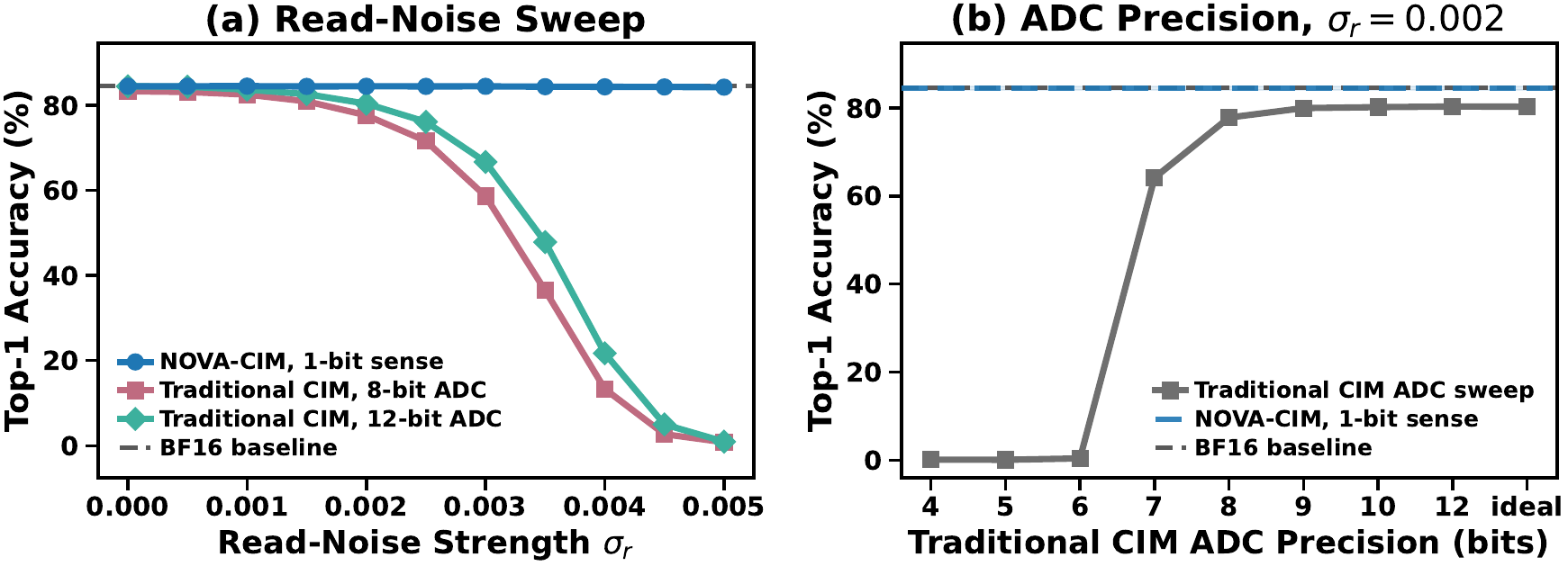}
\caption{ViT-Base evaluation on ImageNet-1k. (a) Top-1 accuracy versus read noise strength. (b) Top-1 accuracy versus ADC precision at $\sigma_r=0.002$.}
\label{fig:e004_read_noise_comparison}
\end{figure}

\textbf{Read Noise Robustness.}
Section~\ref{sec:dynamic_read_noise_averaging} predicts that zero-mean read noise that is independent across cycles leaves the expected output unchanged and is suppressed by temporal averaging. Consistently, Figure~\ref{fig:e004_read_noise_comparison}(a) shows that NOVA-CIM retains $84.48\%$ Top-1 accuracy at $\sigma_r=0.002$, close to the $84.51\%$ BF16 baseline, while ADC-CIM degrades. Figure~\ref{fig:e004_read_noise_comparison}(b) shows that increasing ADC precision saturates near $80.3\%$, because converter precision cannot remove noise already present in the analog current.

\begin{figure}[!t]
\centering
\includegraphics[width=\columnwidth]{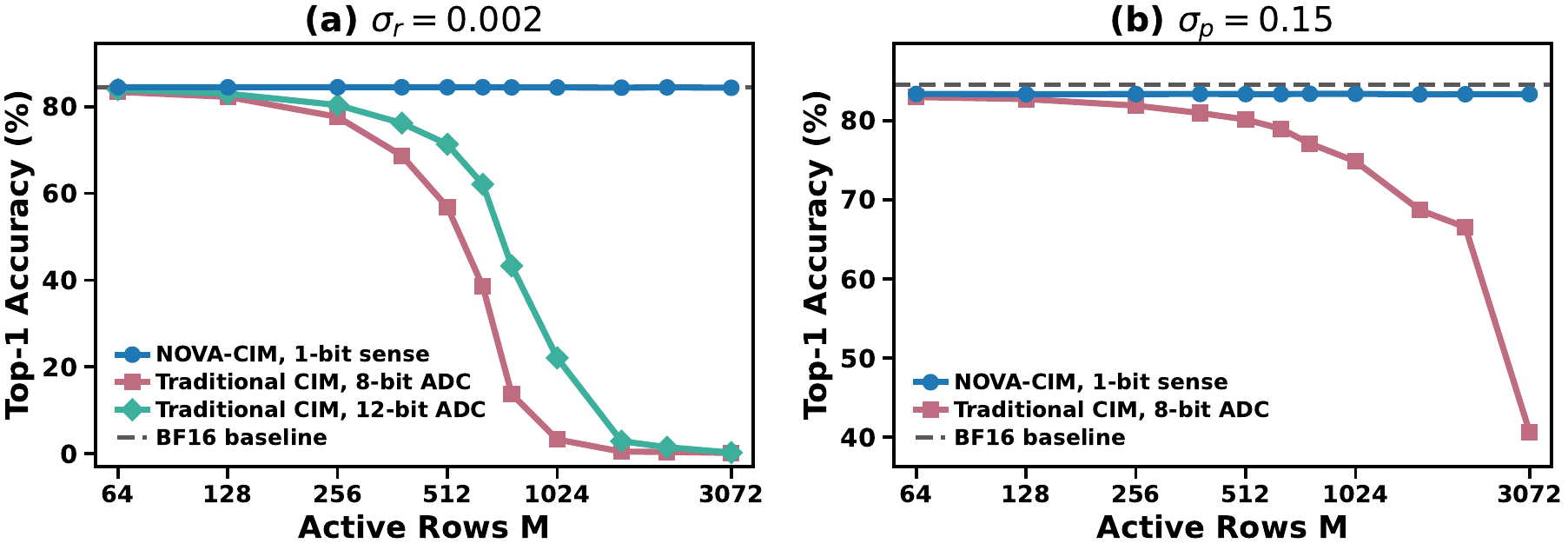}
\caption{ViT-Base active row sweep on ImageNet-1k. (a) Read noise with $\sigma_r=0.002$. (b) Programming noise with $\sigma_p=0.15$.}
\label{fig:e006_active_rows_sensitivity}
\end{figure}

\textbf{Active Row Scalability.}
Figure~\ref{fig:e006_active_rows_sensitivity} examines whether the noise robustness of NOVA-CIM persists as the number of active rows increases. Under $\sigma_r=0.002$, Figure~\ref{fig:e006_active_rows_sensitivity}(a) shows that NOVA-CIM maintains approximately $84.4\%$ Top-1 accuracy from $M=64$ to $3072$, whereas both ADC-CIM configurations fall below $0.3\%$ at $M=3072$. Under $\sigma_p=0.15$, programming noise reduces NOVA-CIM accuracy to approximately $83.35\%$, but this degradation remains nearly constant as $M$ increases. In contrast, ADC-CIM becomes increasingly sensitive to both noise sources as $M$ grows; for example, its 8-bit accuracy under programming noise decreases from $82.97\%$ to $40.67\%$. Thus, NOVA-CIM suppresses dynamic read noise and prevents the effect of programming noise from being amplified as the number of active rows increases.

\begin{figure}[!t]
\centering
\includegraphics[width=\columnwidth]{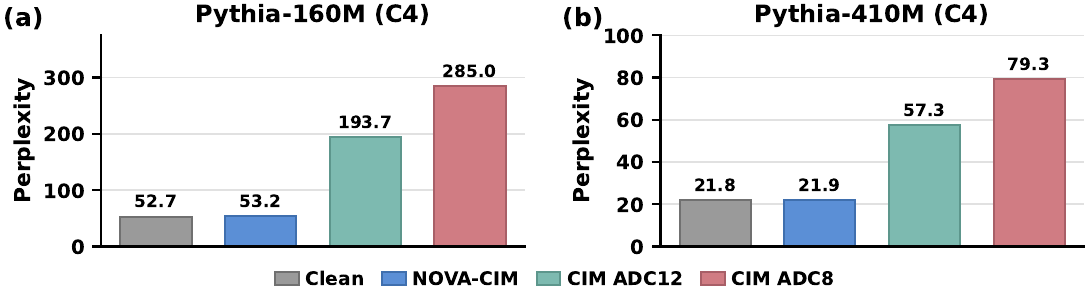}
\caption{Perplexity on the C4 validation set for Pythia-160M and Pythia-410M under clean, NOVA-CIM, 8-bit ADC-CIM, and 12-bit ADC-CIM configurations.}
\label{fig:e007_generation_ppl}
\end{figure}

\textbf{Generalization to Language Models.}
Figure~\ref{fig:e007_generation_ppl} evaluates Pythia-160M and Pythia-410M on C4 under $\sigma_r=0.002$. NOVA-CIM keeps perplexity within $1\%$ of the clean baseline for both model sizes, whereas even 12-bit ADC-CIM increases perplexity by $2.6$--$3.7\times$. This result shows that the tolerance of NOVA-CIM to read noise extends from image classification to autoregressive language modeling.

\textbf{Implementation Cost.}
Table~\ref{tab:cim_interface} compares ADC-CIM and NOVA-CIM interfaces at the same 65\,nm process node. Readout area drops from $3.88{\times}10^{-4}$\,mm$^2$ for an 8.5-bit time-domain ADC (TD-ADC) to $1.36{\times}10^{-4}$\,mm$^2$ for a TIA and latched SA, a $2.84\times$ reduction. Encoding area drops from $4.00{\times}10^{-4}$\,mm$^2$ for a word-line input DAC to $3.15{\times}10^{-4}$\,mm$^2$ for an LFSR SNG, a $1.27\times$ reduction. The larger gain is on readout, where one-bit sensing replaces multi-bit digitization.

Table~\ref{tab:fpga_array_util} gives post-route utilization of a $64{\times}128$ design on XCVU37P. The array uses $329775$ LUTs. Each column MAC uses $2569$ LUTs with $65$ LUTs on the random-reference path, and the shared input SNG uses $2939$ LUTs. Combined with Table~\ref{tab:sng_reuse}, clustered encoding further reduces encoder LUT cost while keeping MAC bias near zero.

\begin{table}[!t]
\centering
\caption{Per-channel interface area of ADC-CIM and NOVA-CIM at the same 65\,nm process node.}
\label{tab:cim_interface}
\renewcommand{\arraystretch}{1.15}
\setlength{\tabcolsep}{4pt}
\small
\begin{tabular}{@{}lcccc@{}}
\toprule
& \multicolumn{2}{c}{\textbf{Encoding}} & \multicolumn{2}{c}{\textbf{Readout}} \\
\cmidrule(lr){2-3}\cmidrule(lr){4-5}
\textbf{Interface}
& Area (mm$^2$) & Impl.
& Area (mm$^2$) & Impl. \\
\midrule
ADC-CIM
& $4.00{\times}10^{-4}$ & WL input DAC
& $3.88{\times}10^{-4}$ & 8.5-bit TD-ADC \\
NOVA-CIM
& $3.15{\times}10^{-4}$ & LFSR SNG
& $1.36{\times}10^{-4}$ & TIA + latch SA \\
\midrule
Reduction
& $1.27\times$ & ---
& $2.84\times$ & --- \\
\bottomrule
\end{tabular}
\end{table}

\begin{table}[!t]
\centering
\caption{FPGA post-route utilization of the $64{\times}128$ digital datapath on XCVU37P.}
\label{tab:fpga_array_util}
\renewcommand{\arraystretch}{1.12}
\begin{tabular}{@{}lrrrr@{}}
\toprule
\textbf{Module} & \textbf{Inst.} & \textbf{LUT} & \textbf{FF} & \textbf{DSP} \\
\midrule
\multicolumn{5}{l}{\textit{Per-column MAC core}} \\
\quad Reference path & 1 & 65 & 66 & 2 \\
\quad Compute core & 1 & 2510 & 1578 & 4 \\
\textbf{Column MAC} & 1 & \textbf{2569} & \textbf{1660} & \textbf{6} \\
\midrule
\multicolumn{5}{l}{\textit{Array}} \\
\quad Input SNG & 64 & 2939 & 2048 & 0 \\
\quad Column MAC & 128 & 325004 & 210432 & 768 \\
\textbf{NOVA-CIM array} &  & \textbf{328943} & \textbf{212480} & \textbf{768} \\
\midrule
\textbf{Total} &  & \textbf{329775} & \textbf{212480} & \textbf{768} \\
Device &  & 1303680 & 2607360 & 9024 \\
Utilization &  & 25.30\% & 8.15\% & 8.51\% \\
\bottomrule
\end{tabular}
\end{table}

\section{Conclusion}

This paper presented NOVA-CIM, a probability-domain CIM readout using random-reference one-bit sensing and temporal counting, yielding an unbiased normalized MAC estimator with variance \(O(1/N)\). Zero-mean dynamic read noise causes no first-order bias and is suppressed by \(1/N\) averaging, while spatial input correlation preserves the MAC expectation and only increases instantaneous current variance. ViT-Base evaluation shows \(84.48\%\) Top-1 accuracy under read noise, close to the \(84.51\%\) BF16 baseline, and SNG reuse keeps MAC bias near zero with RMSE and SCC increasing as predicted.

\clearpage
\begin{acks}
This project is supported in part by the Theme-based Research Scheme (TRS) projects T45-701/22-R and T46-415/25-R, and GRF Project 17203224 of the Research Grants Council (RGC), Hong Kong SAR, and in part by the AVNET-HKU Emerging Microelectronics \& Ubiquitous Systems (EMUS) Lab.
\end{acks}

\bibliographystyle{ACM-Reference-Format}
\bibliography{bibliography}

\end{document}